\documentstyle[12pt]{article}
\begin{document}

\title{Global Bethe lattice consideration of the \\ spin-1 Ising model}
\author{A. Z. Akheyan\thanks{e-mail: akheyan@vxc.yerphi.am} \
and N. S. Ananikian\thanks{e-mail: ananikian@vxc.yerphi.am} \vspace{1 cm}\\
Yerevan Physics Institute,\\
{\small Alikhanian Br. str. 2, 375036, Yerevan, Armenia}}

\date{}

\maketitle


\begin{abstract}
The spin-1 Ising model with bilinear and biquadratic exchange interactions
and single-ion crystal field is solved on the Bethe lattice using
exact recursion equations. The general procedure of critical properties
investigation is discussed and full set of phase diagrams are constructed for
both positive and negative biquadratic couplings. A comparison with the
results of other approximation schemes is done.
\end{abstract}

\newpage

\section {Introduction.}

The spin-1 Ising model with most general up-down symmetry, known
also as a Blume-Emery-Griffiths (BEG) model has attracted recently a great
attention as a simple model with rich and interesting phase structure.
It was originally
introduced \cite{B-E-G} in order to explain the phase separation and
superfluidity in
the $^3$He-$^4$He mixtures and latter has been developed to describe
another multi-component physical systems, such as methamagnets,
liquid crystal mixtures, microemulsions, semiconductor allows, etc.

The model is defined by Hamiltonian:
\begin{equation}
\label{ham}-\beta {\cal H}=J\sum\limits_{<ij>}s_is_j+K\sum%
\limits_{<ij>}s_i^2s_j^2-\Delta \sum\limits_is_i^2
\end{equation}
where $s_i$ takes the values $\pm1,\,0$ at each lattice site, $\langle
ij\rangle $ denotes a summation over all nearest-neighbour pairs, $J$ and $
K$ are correspondingly the bilinear and biquadratic interaction constants, $%
\Delta$ \ is a single-ion crystal field.

The spin-1 Ising model was solved exactly only on
two-dimensional honeycomb lattice in a subspace of interacting constants
$e^K\cosh J=1$ \cite{ES-1,ES-2,ES-3} (recently such solution has been
found for higher spin-S models as well \cite{Suz,S3/2,H}) but
its critical properties for positive $J, K > 0$
were well established by different approximation techniques
\cite{B-E-G,MFA-1,MFA-2,RG-1}.
The most interesting result here was the first occurrence (in theoretical
model) of the tricritical point, at which the second
order phase transition line ($\lambda$-line) turns to the first order one.

The picture is practically not changed for negative bilinear couplings $J<0$.
On the bipartite lattice (i.e. the lattice which can be divided on two
sublattices A and B, such that every site belonging to A is surrounded only
by sites belonging to B and vice versa) the region $J<0$ is mapped on
the region $J>0$ by redefining the spin directions on one sublattice. As a
result we obtain the same phase diagrams, where only ferromagnetic
phase is replaced by antiferromagnetic one. Hence with no loss of
generality we can consider only the case $J>0$.

On the other hand, the negative values of biquadratic coupling K change
the situation drastically. The region $K/J<0$ is now a subject of
intensive studying. A new staggered quadrupollar phase (also called
antiquadrupolar) was predicted and investigated on the square lattice
by means of mean-field
approximation (MFA) and by Monte Carlo (MC) simulations \cite{MC-1}.
The direct first order transition was found from antiquadrupolar (a)
to ferromagnetic (f) phase, but it was not confirmed by the
recent MC \cite{MC-3}
and cluster variation method (CVM) \cite{CVM-1} studies.
Now it seems that $a$ and $f$
phases are always separated in two dimension by disordered phase (d) and
they meet only at $T=0$. This direct $a \leftrightarrow f$ transition
was however established
on three-dimensional cubic lattice \cite{HB-1,HB-2,BA}.
The global MFA analysis \cite{HB-1} on this lattice showed also
a number of other remarkable features, such as doubly reentrant
behaviour at $0>K/J>-1$ and new staggered ferrimagnetic phase which appears
between $a$ and $f$ phases at $K/J<-1$.
Latter investigations \cite{BA,N,NB,CVM-2,CVM-3} mainly confirmed these
results, however a number of contradictions still remain.
This makes interesting the further consideration of the model, especially
by applying another approximation tools.

In the present paper we consider the solution of the spin-1 Ising model
on the Bethe lattice. We review both positive and negative values of
biquadratic coupling $K$ and show that, though in main details we reproduce
the phase diagrams obtained by other authors, there are some essential
differences, concerning the place and order of phase transitions.

The paper is organized as follows. In the Sect. 2 we introduce the model
and derive some analytical expressions including the set of exact recursion
equations. The procedure of critical properties investigation based on
these equations is described in Sect. 3. Resulting phase diagrams are
presented and discussed in Sect. 4. Final Sect. 5. is devoted to Conclusion.

\vspace{0.5 cm}
\section {Model formulation.}

The BEG model is characterized by two order parameters, magnetization $\;m\;$
and qudrupolar moment $\,q$:

\begin{equation}
\label{od}m=\langle s_i\rangle, \;\;\;\;\; q=\langle s_i^2\rangle
\end{equation}

However to account the possible two-sublattice structure we need
actually four order parameters: $m_{A,B}=\langle s_i\rangle _{A,B}$
and $q_{A,B}=\langle s_i^2\rangle _{A,B}$,
where A, B denotes sublattices. These parameters define the four
different phases of the BEG model:

$$
\begin{array}{lll}
\mbox {1. disordered phase } (d): & m_A=m_B=0, & q_A=q_B \\
\mbox {2. ferromagnetic phase } (f): & m_A=m_B\neq 0, & q_A=q_B \\
\mbox {3. antiquadrupolar phase } (a): &  m_A=m_B=0, & q_A\neq q_B \\
\mbox {4. ferrimagnetic phase } (i): &  0\neq m_A\neq m_B\neq 0, & q_A\neq q_B
\end{array}
$$

\vspace{0.2 cm}
The Bethe lattice consideration for any model is based, in some way,
on one or more exact recursion equations. We construct these
equations in the following way \cite{BEG-old}. Taking into
account the shell structure of Bethe lattice (Fig. 1) one can express
the partition
function $Z$ of the model on the finite $n$-shell lattice in the form:
\begin{equation}
\label{Z}Z = \sum\limits_{\{s\}}-\beta{\cal H} =
\sum\limits_{s_0}\exp(-\Delta\,s_0^2){[g_n(s_0)]}^z
\end{equation}
where $z$ is a lattice coordination number,
$s_0$ denotes the central spin and $g_n(s_0)$ is a contribution to the
partition function of one lattice brunch, starting from the central site
with fixed spin value $s_0$. The latter is readily connected with
$g_{n-1}(s_1)$ :
\begin{equation}
\label{g}g_n(s_0)=\sum\limits_{s_1}\exp(Js_0s_1+Ks_0^2s_1^2-\Delta\,s_1^2)
[g_{n-1}(s_1)]^{z-1}
\end{equation}

Introducing new notations:
\begin{equation}
x_n=\frac{g_n(+)}{g_n(0)},\,\,\,\,\,\,\,y_n=\frac{g_n(-)}{g_n(0)}
\end{equation}
and summing up over all values of central spin $s_0$ (i.e. $\pm1,\,0$)
we obtain a set of two recursion equations:
\begin{eqnarray}
\label{recur}
x_{n+1}=\varphi(x_n,y_n) & & y_{n+1}=\varphi(y_n,x_n) \nonumber \\
 & & \\
\mbox{where} &\;\;\;\;\; & \varphi(u,v)=\frac{e^\Delta
+e^K(e^Ju^{z-1}+e^{-J}v^{z-1} )}
{e^\Delta +u^{z-1}+v^{z-1} } \nonumber
\end{eqnarray}

The values $x$ and $y$ have no direct physical sense, but one can express
in terms of $x$ and $y$ all thermodynamic functions of interest. Thus order
parameters (\ref{od}) will be written in the form:

\begin{equation}
\label{m}m=\frac{x^{z}-y^{z}}{e^\Delta
+x^{z}+y^{z}}
\end{equation}

\begin{equation}
\label{q}q=\frac{x^{z}+y^{z}}{e^\Delta
+x^{z}+y^{z}}
\end{equation}

Using equations (\ref{Z}-\ref{recur}) we can writte also the expression
for the free energy:

\begin{equation}
-\beta f=\frac{1}{N}\ln Z
\end{equation}

in the form:

\begin{equation}
\label{f}
-\beta f = \ln \left[ 1+e^{-\Delta }(x^z+y^z)\right] +
\frac{z}{2-z}\ln \left[ 1+e^{-\Delta }(x^{z-1}+y^{z-1})\right]
\end{equation}

\vspace{0.5 cm}
\section{Critical properties investigation.}

The equations (\ref{recur}) form an iteration sequence $\{x_n, y_n\}$,
which in the thermodynamic limit converges to stable fixed points. Via
the expressions (\ref{m}-\ref{f}) these points completely define the
possible states
of the system. The remarkable points of this approach is that
non-staggered phases are described by the single fixed points
$\{x_n, y_n\} \rightarrow \{x, y\}$, while the staggered phases appear
as a 2-cycle doubled points \cite{my-p}:

\[
\{x_n, y_n \} \; \rightarrow \left\{
\begin{array}{ll}
\{x_A, y_A\} & \mbox{for odd n} \\
\{x_B, y_B\} & \mbox{for even n}
\end{array}
\right.
\]

This property can be explained by the fact that all sites of each
individual shell of the Bethe lattice belong to the same sublattice.

Thus, noting from (\ref{m}) that $m=0$ means $x=y$, we can, in our case,
classify the four mentioned phases as follows:

$$
\begin{array}{lll}
\mbox {1. disordered phase }(d): & x=y,\;\;\; & \mbox { single fixed point;}\\
\mbox {2. ferromagnetic phase }(f): & x\neq y, & \mbox{ single fixed point;}\\
\mbox {3. antiquadrupolar phase }(a): & x=y, & \mbox { period doubling;}\\
\mbox{4. ferrimagnetic phase }(i): & x\neq y, & \mbox{ period doubling.}
\end{array}
$$

\vspace{0.2 cm}
It is possible to obtain the full bifurcation picture, including chaos
on some hierarchical lattices \cite{our-chaos}, but not in the bipartite
case (which is the Bethe lattice with nearest-neighbour interactions). We
have only first period doubling and it is well known from the theory of
iteration processes \cite{shust} that all our points of interest
(i.e. stable fixed points and 2-cycle doubled points) can be found
among the solutions of the set of four equations:

\begin{equation}
\label{st}
\left\{
\begin{array}{ll} x_A\,=\,\varphi(x_B,\,y_B,\,J,\,K,\,\Delta) & \mbox{(i)}\\
y_A\,=\,\varphi(y_B,\,x_B,\,J,\,K,\,\Delta) & \mbox{(ii)}\\
x_B\,=\,\varphi(x_A,\,y_A,\,J,\,K,\,\Delta) & \mbox{(iii)}\\
y_B\,=\,\varphi(y_A,\,x_A,\,J,\,K,\,\Delta) & \mbox{(iv)}
\end{array}
\right.
\end{equation}
\vspace{0.2 cm}

The physical stable solutions of this set define the pure states of
the model. As a matter of fact there is no need to solve (\ref{st}) in
general form. Knowing apriori the possible phases of the model
we can separate the solutions of (\ref{st}) concerning the given phase.
Thus the {\it disordered} phase $(x=y,\;A=B)$ can be defined by single
equation, either (i) or (ii). This phase is not degenerate.
For {\it ferromagnetic} phase $(x \neq y,\;A=B)$ it is enough to consider
two equations, namely (i) and (ii) and to exclude the disordered solution.
This phase is doubly-degenerate $(m \leftrightarrow -m)$ due to obvious
symmetry of these two equations
under the $(x \leftrightarrow y)$ transformation. The {\it antiquadrupolar}
phase $(x=y,\;A\neq B)$ again can be found from equations (i) and (iii),
after excluding the disordered solution. This phase is two-fold degenerate
because of the $A\leftrightarrow B$ symmetry. And only for {\it ferrimagnetic}
phase we have to consider all four equations, but they are simplified by
excluding previous solutions. Ferrimagnetic phase is four-fold
degenerate since the $(x \leftrightarrow y)$ and $(A \leftrightarrow B)$
symmetry. Of course, the last two phases are also infinitely degenerate with
non-zero residual entropy.

Further procedure can be roughly described by the following steps:

1) The intersections of the solutions, describing the different phases,
give us, generally speaking, the points of second order transitions.

2) The presence of several simultaneous solutions at the given
$K,\,J,\,\Delta$ again generally speaking means the co-existing phases
and first order transition, which should be located by matching free
energies (\ref{f}) of these phases.

3) The intersections of first and second order critical lines give the
critical and multicritical points of several types.

On Fig. 2 we illustrate this procedure on a simple example of appearance
of the tricritical point. The disordered and ferromagnetic solutions of
(\ref{st}) are shown on a crystal field $\Delta$ versus quadrupolar
moment $q$
plot for different fixed temperatures $1/zJ$. In high temperature region
the $d$ and $f$ solutions intersect in their stable parts and this leads
to a second order transition. At a low temperature the $d$ line does not meet
the stable part of $f$ line and first order transition takes place.
Tricritical point is observed in an intermediate situation.

\vspace{0.5 cm}
\section{Phase diagrams}

The resulting phase diagrams
for spin-1 Ising model are constructed on the Bethe lattice with coordination
number $z=4$. As it is common we plot them as a distinct constant
$K/J$ cross-sections, in the temperature $1/zJ$ versus crystal field
$\Delta/zJ$. We obtain eight qualitatively different diagrams in the whole
range of parameter space (Fig. 3).

The $0>K/J>-1$ counterpart is characterized by absence of doubled points
of the recursion sequence $\{x_n,y_n\}$ and hence represents the non-staggered
region of the BEG model with two phases $d$ and $f$. The first three phase
diagrams (Fig. 3a-c) are well known from the pioneer work of Blume, Emery
and Griffiths \cite{B-E-G}. Our results are quite agree with general
picture. The tricritical point doesn't appear for large positive $K/J$
(Fig. 3a), since the second order line, limiting ferromagnetic phase
from above, terminates earlier at the critical point $E$ by the first
order line limiting $f$
phase from the right. The latter line itself terminates at critical
point $C$, and in higher-temperature segments two subphases of disordered
phase with different dense co-exist.
For the $K/J$ close to 3 (Fig. 3b) we observe
tricritical point $T$ at which second order line ($\lambda$-line) turns to
the first order one and also the triple point $R$, where three different first
order transitions meet. This structure disappears as $K/J$ goes down, and
simple tricritical point is seen (Fig. 3c) for $K/J$ close to 0
(from both positive and negative sides).

Very interesting critical phenomenon, called doubly reentrant behaviour,
takes place for the values $-0.35>K/J>-1$ (Fig. 3d).
At a fixed crystal field $\,\Delta/zJ\,$ the model exhibits the
disorder-ferromagnetic-disorder-ferromagnetic sequence of
phases as temperature is lowered. The dependence of order parameters
versus temperature is shown for this region on Fig. 4.
Doubly reentrant structure shrinks to a zero
temperature and at $K/J=-1$ the only second order line remains,
which reaches the $T=0$ axes at a point $\Delta=0$ (Fig. 3e).

Comparing our results with other approximations, we note the following
differences.
Doubly reentrance appears in our model for lower values of $K/J$
and continue till $K/J=-1$, hence we do not observe
the internal critical point structure of ferromagnetic phase. This
structure was found by MFA \cite{HB-1} and confirmed by RG studies
\cite{NB}, but it was also not established by CVM \cite{CVM-3}. Besides we
would like to mention that nowhere in region $0>K/J>-1$ we obtain the
single reentrant part, though it was found by other authors.

The 2-cycle doubled solutions of (\ref{st}) appear at $K/J<-1$, and this
means the emergence of new phases with broken sublattice symmetry.

At $-3<K/J<-1$ (Fig. 3f) the {\it antiquadrupolar} phase is separated from $d$
phase by second order line and from $f$ phase by first order one. These two
lines meet with the second order line, separating the $d$ and $f$ phases, at
bicritical point $B$. The {\it ferrimagnetic} phase lies at the
low-temperature region and is separated by the second order line from $a$
phase and by first order line from $f$ phase. These two lines meet with
{\it a--f} \/\/ first order line at critical end point $E$ and at point $S$.
The latter point locates at  $T=0,\/\Delta/zJ=K/J+1$ and
describes the macroscopically degenerated ground state with non-zero
residual entropy.

As $K/J$ lowers the point $E$ approaches to the $B$ point and they coincide
at $K/J=-3\,$ in the point $A$ (Fig. 3g), such that there is no
direct transition from $a$ to $f$ phase. Thus $A$ is a new multicritical
point at which three second order lines and one first order line meet.
Note that the first order {\it i--f}\  transitions locate at straight
vertical line and this locus $K/J=-3,\,\Delta /qJ=-2$ corresponds to
zero-field 3-state antiferromagnetic Potts model with Hamiltonian:
\begin{equation}
\label{AFP}-\beta {\cal H}=-2J\sum\limits_{<ij>}\delta
_{s_i},_{s_j},\,\,\,\,\,\,J>0
\end{equation}

At $K/J<-3$ the transitions from $i$ to $f$ phase in high temperature region
become of the second order (Fig. 3h). As a result the new tricritical point
$T^{\prime}\,$ arises inside the ordered region. The multicritical point $A$
turns to the tetracritical point $M$, at which four second order lines meet
with different slopes.

Though in general the last three diagrams are similar to those obtained by
MFA, there is one essential difference: in our approach the transitions
from $a$ to $i$ phase are of second order (first order in MFA), while the
transitions from $i$ to $f$ phase are of first order at
least for $K/J\geq -3$ (always second order in MFA).
This means in particular that we observe another type of ferrimagnetic phase,
which co-exists with $f$ phase and is caused by instability of $a$ phase
against the spontaneous magnetization. Such phase was found by CVM
\cite{CVM-2,CVM-3}, but only in high-temperature region, while it is the only
ferrimagnetic phase present in our consideration. We would like also
to mention the very narrow region of occurrence of ferrimagnetic
phase.

\vspace{0.5 cm}
\section{Conclusion}

Using exact solution on the Bethe lattice we have constructed the full
set of phase diagrams for spin-1 Ising model for both positive and
negative biquadratic coupling $K$. These diagrams feature all recently
found properties of the model, including doubly-reentrant behaviour,
staggered quadrupolar and ferrimagnetic phases and great number of different
critical and multicritical points. Thus we can say about the validity of
Bethe lattice approximation and summaries some advantages of this
method. The quantitative comparison of the results has not been our purpose,
but Bethe lattice solution was shown to be more exact than MFA
\cite{our-j.p.,p-gi1}. Besides it is quite easy to use and it provides the
analytical expression of all thermodynamic functions of interest, so
the complete information about the system under the study can be
obtained.

As to the mentioned disagreements, especially in the staggered region,
they may be caused by dimensionality effects \cite{HB-2}. The Bethe lattice
is effectively infinite dimensional, but one can successfully approximate
the real lattices in different dimensions by changing the coordination
number (this was shown in particular in the global consideration of the
antiferromagnetic Potts model \cite{p-gi2}). We have constructed the phase
diagrams for the Bethe lattice with coordination number $z=4$. We
do not consider the simplest case $z=3$ since it leads to qualitatively
different phase diagrams, similar to those, which were obtained on
two-dimensional lattices (see Introduction). Our preliminary study shows
also that phase diagrams change in case of greater coordination number.
This question we are going to clarify in our further work.

\vspace{0.5 cm}
\section*{Acknowledgements}

This work was partly supported by the grant 211-5291YPI of the German
Bundesministerium f\"ur Forrschung und Technologie, ISF supplementary
grant and grant INTAS-93-633. We wish to thank R. Flume, K.A. Oganessyan,
N.Sh. Izmailian and E.Sh.~Mamasakhlisov for help in work and useful
discussions. We are also grateful to International Atomic Energy Agency,
UNESCO and personally to Prof. S.~Randjbar-Daemi for hospitality
at the International Centre for Theoretical Physics, Trieste, where the
part of this work was done.

\newpage\

\newpage\

\begin{center}
{\bf Figures captions:}
\end{center}
\vspace{0.5cm}
\begin{description}

\item[Fig. 1.] Bethe lattice with coordination number $z=3$
\item[Fig. 2.] Graphical representation of the solutions of equation (\ref{st})
on a $\Delta$ versus $q$ plot at different temperatures: $1/zJ=0.4$ (a),
$1/zJ=0.15$ (b), $1/zJ=0.23$ (c) and constant $K/J=0$. $d$ and $f$
are respectively the
disordered and ferromagnetic phases. Solid lines represent
the stable solutions, dashed lines -- unstable and non-physical
solutions. Thick solid lines show the main behaviour of the system.
\item[Fig. 3.] Phase diagrams of the spin-1 Ising model on the Bethe lattice
with coordination number $z=4$ at constant $K/J$ values: 5 (a), 3 (b),
-0.1 (c), -0.8 (d), -1 (e), -2.5 (f), -3 (g), -3.5 (h). Disordered {\bf d},
ferromagnetic {\bf f}, antiquadrupolar {\bf a} and ferrimagnetic {\bf i}
phases are present. Dashed and solid lines indicate respectively the first-,
second order transitions and $C, E, R, T, T^{\prime} S, B, A, M$ indicate the
critical and multicritical points of different types (see in text).
Some fair details are shown in the insets.
\item[Fig. 4.] The dependence of order parameters $m, q$ versus temperature
$1/zJ$ at a constant $\Delta/zJ=0.099, \; K/J=-0.8\,$ in the doubly
reentrant region. Close, open arrows indicate respectively the places of
first-, second order transitions.

\end{description}

\begin{thebibliography}{99}
\bibitem{B-E-G}  M. Blume, V. J. Emery and R. B. Griffiths, Phys.Rev. A
{\bf 4}, 1071 (1971).

\bibitem{ES-1}  T. Horiguchi, Phys. Lett. A {\bf 113}, 425 (1986).

\bibitem{ES-2}  F. Y. Wu, Phys. Lett. A {\bf 116}, 245, (1986).

\bibitem{ES-3}  A. Rosengren and R. H\"aggkvist, Phys. Rev. Lett.
{\bf 63}, 660 (1989)

\bibitem{Suz}  A. Lipowski and M. Suzuki, Physica A {\bf 193}, 141 (1994)

\bibitem{S3/2}  N. S. Ananikian and N. Sh. Izmailian, Phys. Rev. B {\bf 50},
6829-6832 (1994)

\bibitem{H}  T. Horiguchi, Physica A {\bf 214}, 452 (1995).

\bibitem{MFA-1}  D. Mukamel and M. Blume, Phys. Rev. A {\bf 10}, 619 (1974).

\bibitem{MFA-2}  D. Furman, S. Dattagupta and R. B. Griffiths, Phys.Rev. B
{\bf 15}, 441 (1977).

\bibitem{RG-1}  A. N. Berker and M. Wortis, Phys. Rev. B {\bf 14}, 4946 (1976).

\bibitem{MC-1}  M. Tanaka and T. Kawabe, J. Phys. Soc. Jpn. {\bf 54},
2194, (1985).

\bibitem{MC-3}  Y. L. Wang, F. Lee and J. D. Kimel, Phys. Rev. B {\bf 36},
8945, (1987).

\bibitem{CVM-1}  A. Rosengren and S. Lapinskas, Phys. Rev. B {\bf 47},
2643-2647 (1992).

\bibitem{HB-1}  W. Hoston and A. N. Berker, Phys. Rev. Lett. {\bf 67},
1027-1030 (1991).

\bibitem{HB-2}  W. Hoston and A. N. Berker, J. Appl. Phys. {\bf 70},
6101-6103 (1991).

\bibitem{BA} K. Kasono and I. Ono, Z. Phys. B {\bf 88}, 205-212 (1992).\\
K. Kasono and I. Ono, Z. Phys. B {\bf 88}, 213-214.

\bibitem{N}  R. R. Netz, Europhys. Lett. {\bf 17} (4), 373-377 (1992).

\bibitem{NB}  R.R. Netz and A. N. Berker, Phys. Rev B {\bf 47}, 15019-15022
(1993).

\bibitem{CVM-2}  A. Rosengren and S. Lapinskas, Phys. Rev. Lett {\bf 71},
165-168 (1993).

\bibitem{CVM-3}  S. Lapinskas and A. Rosengren, Phys. Rev. B {\bf 49},
15190-15196 (1994).

\bibitem{BEG-old}  A. R. Avakian, N. S. Ananikian and N. Sh. Izmailian,
Phys. Lett. A {\bf 150}, 163-165 (1990).\\
N. S. Ananikian, A. R. Avakian and N. Sh. Izmailian,
Physica A {\bf 172}, 391-404 (1991).

\bibitem{my-p}  A. Z. Akheyan and N. S. Ananikian, Phys. Lett. A {\bf 186},
171-174 (1994).\\
A. Z. Akheyan and N. S. Ananikian, JETP (Russia) {\bf 107},
196-208 (1995).

\bibitem{our-chaos}  N. S. Ananikian, R. Lusiniants and K. A. Oganessyan,
JETP Lett. (Russia) {\bf 61}, 482-486 (1995).

\bibitem{shust} see for exapmple H. G. Schuster, Deterministic chaos
(physik-Verlag, Weinheim, 1984).

\bibitem{our-j.p.}  A. Z. Akheyan and N. S. Ananikian, J. Phys. A {\bf 25},
3111 (1992).

\bibitem{p-gi1}  F. Peruggi, F di Liberto and G. Monroy, J. Phys. A {\bf 16},
811-828 (1983),

\bibitem{p-gi2}  F. Peruggi, F di Liberto and G. Monroy, Physica A {\bf 141},
151-186 (1987).

\end{thebibliography}
\end{document}